\documentclass[10pt,twoside,twocolumn,english,aps,manuscript,aps,preprint,showpacs,superscriptaddress,showkeys]{revtex4}
\usepackage{lmodern}
\usepackage[T1]{fontenc}
\usepackage[latin9]{inputenc}
\usepackage[active]{srcltx}
\usepackage{graphicx}
\usepackage{setspace}

\makeatletter

\newcommand*\LyXThinSpace{\,\hspace{0pt}}

\@ifundefined{textcolor}{}
{%
 \definecolor{BLACK}{gray}{0}
 \definecolor{WHITE}{gray}{1}
 \definecolor{RED}{rgb}{1,0,0}
 \definecolor{GREEN}{rgb}{0,1,0}
 \definecolor{BLUE}{rgb}{0,0,1}
 \definecolor{CYAN}{cmyk}{1,0,0,0}
 \definecolor{MAGENTA}{cmyk}{0,1,0,0}
 \definecolor{YELLOW}{cmyk}{0,0,1,0}
}

\makeatother

\usepackage{babel}
\begin{document}

\title{Paramagnetic magnetization signals and curious metastable behaviour
in field-cooled magnetization of a single crystal of superconductor
2H-NbSe$_{2}$}

\author{Santosh Kumar}

\email{santoshkumar@phy.iitb.ac.in}

\address{Department of Physics, Indian Institute of Technology Bombay, Mumbai
400076, India.}

\author{C. V. Tomy}

\address{Department of Physics, Indian Institute of Technology Bombay, Mumbai
400076, India.}

\author{G. Balakrishnan}

\address{Department of Physics, University of Warwick, Coventry CV4 7AL, UK.}

\author{D. McK Paul}

\address{Department of Physics, University of Warwick, Coventry CV4 7AL, UK.}

\author{A. K. Grover}

\email{arunkgrover@gmail.com}

\address{Department of Condensed Matter Physics and Materials Science, Tata
Institute of Fundamental Research, Mumbai 400005, India.}

\address{Department of Physics, Panjab University, Chandigarh 160014, India.}
\begin{abstract}
We present here some newer characteristics pertaining to paramagnetic
Meissner effect like response in a single crystal of the low $T_{c}$
superconducting compound 2H-NbSe$_{2}$ via a detailed study of effects
of perturbation on the field-cooled magnetization response. In the
temperature range, where an anomalous paramagnetic magnetization occurs,
the field-cooled magnetization response is found to be highly metastable:
it displays a curious tendency to switch randomly from a given paramagnetic
value to a diamagnetic or to a different paramagnetic value, when
the system is perturbed by an impulse of an externally applied ac
field. The new facets revealed in a single crystal of 2H-NbSe$_{2}$
surprisingly bear a marked resemblance with the characteristics of
magnetization behaviour anticipated for the giant vortex states with
multiple flux quanta ($L\Phi_{0}$, $\Phi_{0}=hc/2e$, $L>1$) predicted
to occur in mesoscopic-sized superconducting specimen and possible
transitions amongst such states.
\end{abstract}



\maketitle

\section{Introduction}

The observation of an anomalous paramagnetic magnetization signal
(viz., paramagnetic Meissner effect (PME)), instead of the usual diamagnetic
behaviour, on field\,-\,cooling a superconducting specimen, continues
to attract \cite{key-1,key-2} attention ever since its discovery
\cite{key-3} in ceramic sample(s) of a high $T_{c}$ cuprate. In
addition to high $T_{c}$ cuprates \cite{key-3,key-4,key-5,key-6,key-7,key-8,key-9,key-10,key-11,key-12},
the PME-like attribute is known to occur in a wide variety of other
superconductors \cite{key-13,key-14,key-15,key-16,key-17,key-18,key-19,key-20,key-21,key-22,key-23,key-24,key-25,key-26}.
Numerous explanations (see Ref.\,\cite{key-27} for a review), e.g.,
the d\,-\,wave superconductivity \cite{key-9}, orbital glass \cite{key-27,key-28},
presence of $\pi$ contacts leading to spontaneous currents \cite{key-4,key-8,key-29},
Josephson junctions \cite{key-30}, etc., have been advanced as the
origin of PME like signal(s) in high $T_{c}$ superconductors. A model
proposed by Koshelev and Larkin \cite{key-31} envisaged the possibility
of trapping of magnetic flux in the interior of a superconducting
sample due to inhomogeneous cooling. In such a circumstance, the PME
can occur due to unbalancing between two oppositely directed currents,
viz., (i) a (paramagnetic) current set up in the interior that attempts
to screen the effect of trapped flux and, (ii) a usual (diamagnetic)
current flowing on the surface that shields the external field \cite{key-31}.
However, within the Ginzburg\,-\,Landau framework, the theoretical
works by Moshchalkov \textit{et} \textit{al.} \cite{key-32} and Zharkov
\cite{key-33}, for model cases of mesoscopic cylindrical shaped superconductors,
lead to the proposition that the multi\,-\,quanta vortex matter,
i.e., vortices with multiple flux quanta, $L\Phi_{0}$,~$\Phi_{0}=hc/2e$,~$L>1$,
nucleating below a third critical field ($H_{c3}$), at the onset
of surface superconductivity \cite{key-34} can also give rise to
the PME. 

Recently, a controlled switching of the PME into the usual (diamagnetic)
Meissner effect has been vividly demonstrated by Xing \textit{et al.
}\cite{key-26} in a superconducting ferromagnet Pb-Co nanocomposite,
wherein the source of PME is argued to be related to some different
mechanism other than those stated above. This manipulation of PME
is possible by the change in orientation of magnetic moments of Co
nanoparticles with respect to the external magnetic field \cite{key-26}.
Such a tuning of the anomalous PME and the non-anomalous (diamagnetic)
Meissner effect has rarely been reported in the literature. As another
case of manipulating a PME like response, we have explored some novel
facets of the PME in a single crystal of 2H-NbSe$_{2}$ \cite{key-35},
an example of a conventional low $T_{c}$ superconducting system often
studied to explore novel notions in vortex matter. The results being
reported ahead in 2H-NbSe$_{2}$ present an advancement over studies
earlier reported by some of us in Ca$_{3}$Rh$_{4}$Sn$_{13}$ \cite{key-24}
and Nb \cite{key-17}. The curious behaviour pertains to an unusual
paramagnetic magnetization (\textit{a la }PME) and its manipulation
just below $T_{c}$ in the (isofield) temperature-dependent dc magnetization
($M(T)$) data. We find that in a short window of temperature, where
the PME occurs below $T_{c}$ in 2H-NbSe$_{2}$, an external perturbation
can randomly change the magnetization from a given paramagnetic value
to a larger or lesser paramagnetic value or from a given paramagnetic
state to a diamagnetic one. Across the same temperature interval,
the system can also transit from a given diamagnetic state into a
paramagnetic one after being perturbed by an ac field impulse. The
vortex matter in the field-temperature domain of the PME in 2H-NbSe$_{2}$
is thus found to be highly metastable. The results in a bulk 2H-NbSe$_{2}$
specimen unexpectedly appear to echo the consequences of nucleation
of several metastable multiple flux quanta states ($L>1$) proposed
in the context of mesoscopic samples \cite{key-32,key-33,key-36,key-37,key-38,key-39,key-40}.

\begin{figure}[t]
\begin{centering}
\includegraphics[scale=0.37]{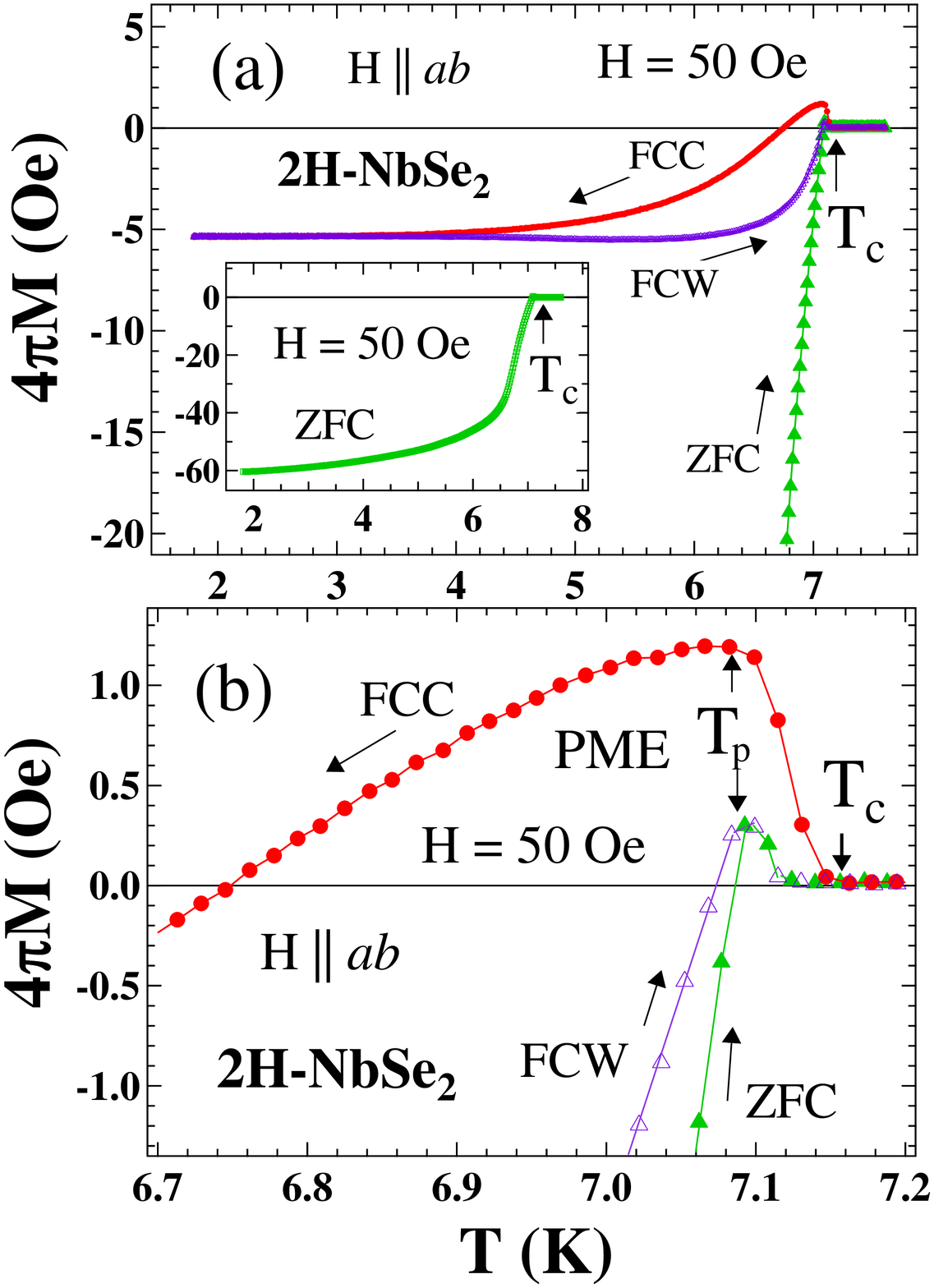}
\par\end{centering}

\protect\caption{{\small{}(Color online) (a) Temperature variation of $M_{ZFC}$, $M_{FCC}$
and $M_{FCW}$ in a field of $H$\, ($\parallel ab$)\,$=50$\,Oe.
The inset in panel (a) illustrates $M_{ZFC}\thinspace(T)$ plot in
the range, $1.8$\,K\,$<T<T_{c}$. Panel (b) elucidates $M(T)$
plots on the expanded scale in the vicinity of $T_{c}$. Anomalous
paramagnetic signals are evident in all three modes. The paramagnetic
magnetization reaches its extreme limit at $T_{p}$ during the field-cooled
cool-down run, $M_{FCC}\thinspace(T)$}.}
\end{figure}

\section{Experimental details}

The single crystal of hexagonal 2H-NbSe$_{2}$ chosen for the present
study is platelet-shaped with a planar area of 4.12\,mm$^{2}$, thickness
of 0.17\,mm and mass of 4.64\,mg. The superconducting transition
temperature ($T_{c}$) for this crystal is measured to be about 7.15\,K.
The crystallographic \textit{ab} plane of the hexagonal 2H-NbSe$_{2}$
coincides with the plane of the platelet, while the direction normal
to it corresponds to the crystallographic \textit{c}\,-\,axis of
2H-NbSe$_{2}$. The magnetization measurements were performed using
a Superconducting Quantum Interference Device\,-\,Vibrating Sample
Magnetometer (SQUID\,-\,VSM, Quantum\,-\,Design, Inc., USA) in
such a way that the magnetic field was directed nominally parallel
to the plane of the platelet (i.e., $H$\,||\,$ab$), with less
than 5 degree error in the angle of alignment. An advantage of investigating
the sample along this orientation is that the demagnetization factor
is small ($\sim10^{-1}$) and the associated geometrical/boundary
effects are expected to be minimal. We estimated, using the flux gate
option of the SQUID\,-\,VSM, the field-inhomogeneity of the superconducting
magnet over a scan length of 8\,mm. It is found to be only of the
order $10^{-2}$\,Oe. During the magnetization measurements, the
amplitude of vibration of the sample was kept small ($\approx0.5$\,mm)
in most of the runs (unless specifically stated otherwise) so that
the possible artefact \cite{key-41} due to field inhomogeneity along
the scan length of the superconducting sample in SQUID\,-\,VSM is
so minuscule that it is of no significance. The effect of changing
the amplitude of sample vibration on the magnetization value (at a
given $H$,\,$T$ value) in the superconducting state of a standard
Indium sample has also been investigated. We registered no difference
in the magnetization response of Indium specimen in its superconducting
state, when the amplitude of vibration in VSM measurements was changed
from 0.5 to 8\,mm. Hence, under the normal circumstances, it can
be stated that the change in amplitude of sample vibration (upto a
maximum of 8\,mm in the present VSM instrument) is not expected to
alter the dc magnetization value of a superconducting sample at a
given $H$ and $T$. In the case of 2H-NbSe$_{2}$ crystal, we have
observed a wide variation in (isofield) magnetization response, ranging
from paramagnetic to diamagnetic values at different amplitudes (see
Fig.~7), which is argued to be a consequence of metastability effects
prevailing in this system, a key focus of the present report.

\section{Results}

\subsection{Paramagnetic signal on field\,-\,cooling}

Figure~1 displays the isofield magnetization responses obtained in
the zero-field cooled ($M_{ZFC}\thinspace(T)$), field-cooled cool-down
($M_{FCC}\thinspace(T)$) and field-cooled warm-up ($M_{FCW}\thinspace(T)$)
modes in 2H-NbSe$_{2}$ in a field of 50\,Oe. For $M_{ZFC}\thinspace(T)$
run, the crystal was initially cooled down to 1.8\,K in nominal zero
field. A field, $H=50$\,Oe, was then applied (such that $H\parallel ab$)
and the magnetization obtained while warming to higher temperatures,
yielding $M_{ZFC}\thinspace(T)$. Without changing the magnetic field,
the $M(T)$ data were again obtained while cooling the sample back
to the lowest temperature ($M_{FCC}\thinspace(T)$). Thereafter, the
magnetization values were measured again while warming it to yield
$M_{FCW}\thinspace(T)$. The $M_{ZFC}\thinspace(T)$ curve shows the
usual fall in diamagnetic response (cf. inset panel of Fig.~1(a))
of a superconducting specimen as the temperature is swept up. 

An expanded portion of $M(T)$ plots of Fig.~1(a), in the proximity
of $T_{c}$ is displayed in Fig.~1(b). Unusual paramagnetic responses
below $T_{c}$ can be observed in all the three modes. It is pertinent
to note that the paramagnetic signals in $M_{ZFC}\thinspace(T)$ and
$M_{FCW}\thinspace(T)$ runs are restricted to a very narrow temperature
interval ($\sim50$\,mK) a little below $T_{c}$. During $M_{FCC}\thinspace(T)$
run, the paramagnetic response nucleating at $T_{c}$ enhances as
the temperature is lowered towards a temperature marked as $T_{p}$.
Thereafter, the $M_{FCC}\thinspace(T)$ values start to decrease,
and cross over to the diamagnetic values at a temperature of about
$6.75$\,K. The diamagnetic $M_{FCC}\thinspace(T)$ signal reaches
a saturated value below $4$\,K (cf. main panel of Fig.~1(a)). During
warm-up run, saturated $M_{FCW}\thinspace(T)$ values remain sustained
upto about $6$\,K. Above this temperature, the diamagnetic $M_{FCW}\thinspace(T)$
values decrease, and they crossover to the paramagnetic values to
merge into the $M_{ZFC}\thinspace(T)$ curve at about $7.08$\,K,
which is nearly the same temperature (marked as $T_{p}$) below which
paramagnetic signal starts to decrease during the corresponding $M_{FCC}\thinspace(T)$
run. It can be noted in Fig.~1(b), that the paramagnetic response
in $M_{FCC}\thinspace(T)$ run is much larger than that in $M_{ZFC}\thinspace(T)$
run or $M_{FCW}\thinspace(T)$ run. This observation implies that
the positive magnetization signal below $T_{c}$ is multi-valued (and
thus metastable) and depends significantly on the thermomagnetic history
of the sample.
\begin{figure}[b]
\begin{centering}
\includegraphics[scale=0.35]{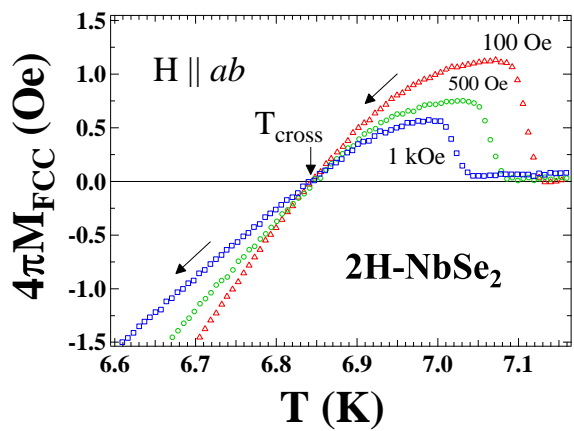}
\par\end{centering}

\begin{singlespace}
\protect\caption{(Color online) Portions of the $M_{FCC}\thinspace(T)$ curves at $H=100$\,Oe,
$500$\,Oe and $1$\,kOe, intersecting unusually at a characteristic
temperature $T_{cross}$.}
\end{singlespace}
\end{figure}

Figure~2 illustrates the $M_{FCC}\thinspace(T)$ curves for $H$\,($\parallel ab$)\,$=100$\,Oe,
$500$\,Oe and $1$\,kOe, which intersect unexpectedly at a temperature,
marked as $T_{cross}$\,($\approx6.84$\,K), where the magnetization
incidentally also crosses the $M=0$ axis. Below $T_{cross}$, the
magnetic behaviour displayed in Fig.~2 is such that the diamagnetic
response decreases as the field increases (($\triangle M/\triangle H$)\,$>0$),
which is typical of a vortex state of a type\,-\,II superconductor
in the field interval, $H_{c1}<H<H_{c2}$. On the other hand, the
magnetization response curves above $T_{cross}$ imply that the paramagnetism
decreases as the applied field increases {[}32{]}.

\begin{figure}[t]
\begin{centering}
\includegraphics[scale=0.32]{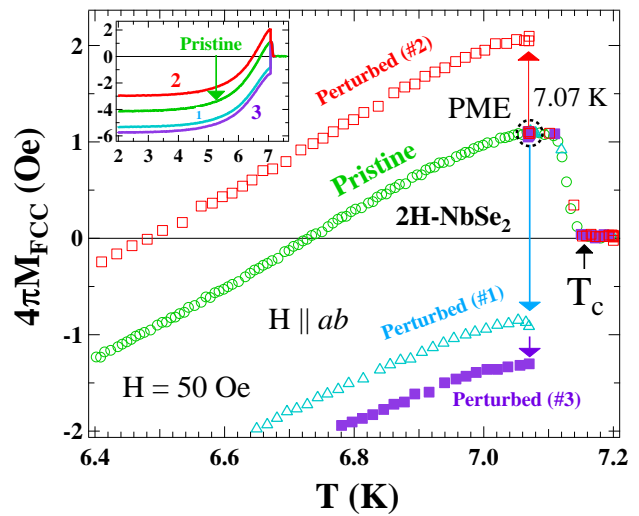}
\par\end{centering}

\begin{singlespace}
\protect\caption{(Color online) $M_{FCC}\thinspace(T)$ data recorded at $H=50$\,Oe
in several runs. In each of these runs, an impulse of an ac field
(peak amplitude\,$=10$\,Oe, frequency\,$=211$\,Hz) was applied
at $T=7.07$\,K (encircled), after which the field-cooled magnetization
change randomly to either paramagnetic or diamagnetic values, and
$M_{FCC}\thinspace(T)$ traverses different paths as displayed for
three different runs, viz., perturbed (\#1), perturbed (\#2) and perturbed
(\#3). The pristine $M_{FCC}\thinspace(T)$ (open circles) was recorded
without perturbing the system throughout the cooling. The inset displays
$M_{FCC}\thinspace(T)$ for pristine and perturbed (\#1, \#2, and
\#3) cases in the full range, $2$\,K\,$<T<7.5$\,K.}
\end{singlespace}
\end{figure}

\subsection{Metastability and non\,-\,uniqueness in magnetization response}

The multi\,-\,valuedness in $M(T)$ observed in Fig.~1(b) motivated
further exploration of magnetization response in 2H-NbSe$_{2}$. We
illustrate in the main panel of Fig.~3, portions of $M_{FCC}\thinspace(T)$
curves in the vicinity of $T_{c}$ recorded at $H=50$\,Oe during
several runs where a perturbation had been induced at an intermediate
stage. First, we recorded the magnetization data (see open circles
in Fig.~3) while field-cooling the sample to $2$\,K, similar to
the way $M_{FCC}\thinspace(T)$ data of Fig.~1 were obtained. This
curve will henceforth be called as \textquotedblleft pristine\textquotedblright{}
in the discussion ahead. We then recorded the $M_{FCC}\thinspace(T)$
data (without changing the external field) while cooling the sample
down to (near) peak temperature ($T_{p}$) of the PME, i.e., $T\approx7.07$\,K,
which is encircled in the main panel of Fig.~3. At this temperature,
the $M_{FCC}\thinspace(T)$ measurements were paused and the system
was momentarily subjected to a perturbation in the form of an ac field
impulse of peak amplitude $10$\,Oe (frequency of ac field\,$=211$\,Hz)
imposed for a duration of $6$\,seconds. It was checked that the
results being reported in Fig.~3 do not depend on the duration of
the impulse. The $M_{FCC}\thinspace(T)$ data recorded after the impulse
treatment showed an abrupt switching from paramagnetic to diamagnetic
value, which is indicated by an arrow pointing towards open triangles
in main panel of Fig.~3. The $M_{FCC}\thinspace(T)$ curve (open
triangles labeled as \textquotedblleft perturbed (\#1)\textquotedblright )
thereafter traverses a path which is akin to the one that seems to
display the usual (diamagnetic) Meissner effect in contrast to the
anomalous PME observed for the pristine curve (open circles). In the
next run, we again cooled the sample down to the same temperature,
$T_{p}$\,($\approx7.07$\,K), and once again applied the same impulse.
Surprisingly, the $M_{FCC}\thinspace(T)$ data now displayed (see
perturbed (\#2) in Fig.~3) an enhancement in the paramagnetic signal
(shown by upward arrow pointing to open squares in Fig.~3). This
is in sharp contrast to the impact of an impulse in the \textquotedblleft perturbed
(\#1)\textquotedblright{} case, which involved switching of the magnetization
from paramagnetic to a diamagnetic value. On further lowering of temperature,
the $M_{FCC}\thinspace(T)$ curve for perturbed (\#2) can be seen
to crossover to diamagnetic values at a temperature ($\approx6.48$\,K),
which is lower than the corresponding crossover temperature of the
pristine $M_{FCC}\thinspace(T)$ curve. In the identical imposition
of an ac impulse in the third run, the $M_{FCC}\thinspace(T)$ data
(perturbed (\#3)) can be switched from paramagnetic to a diamagnetic
value which is even larger than that during perturbed (\#1). On progressively
lowering the temperature below $T_{p}$, the $M_{FCC}\thinspace(T)$
curve for \textquotedblleft perturbed (\#3)\textquotedblright{} (closed
squares) is seen to stay more diamagnetic (at a given $T$) than that
observed in the \textquotedblleft perturbed (\#1)\textquotedblright ,
as shown in Fig.~3.

In the inset panel of Fig.~3, we show the $M_{FCC}\thinspace(T)$
curves for the pristine and perturbed states (\#1, \#2 and \#3) in
the entire temperature range of investigation, i.e., $2$\,K\,$<T<7.5$\,K.
It is curious to note that the saturated $4\pi M$ value at $2$\,K
in a given perturbed run roughly amounts to the summation of the saturated
$4\pi M$ for the pristine curve and the change in magnetization,
$\triangle$\,($4\pi M$) induced at $T_{p}$ due to an impulse treatment.

Subsequent efforts of applying the impulse not only at $T_{p}$ but
also in its vicinity, viz., across $6.75$\,K$<T<T_{c}$ (all data
not shown here), lead to the inference that the $M_{FCC}\thinspace(T)$
value, after perturbation in the PME region, can unpredictably change
very widely from paramagnetic to diamagnetic or from a given paramagnetic
to a less/more paramagnetic value. We also applied impulse with different
amplitudes and at different frequencies of the ac field, however,
the results turned out to be same. In all, we can surmise that the
vortex matter in the temperature regime of PME in 2H-NbSe$_{2}$ is
highly metastable and the magnetization associated with such metastable
states can show a rich diversity, as is evidenced in Fig.~3. When
subjected to a perturbation, the system can transit between these
metastable states, as is reflected from abrupt change(s) in magnetization.
\begin{figure}[t]
\begin{centering}
\includegraphics[scale=0.4]{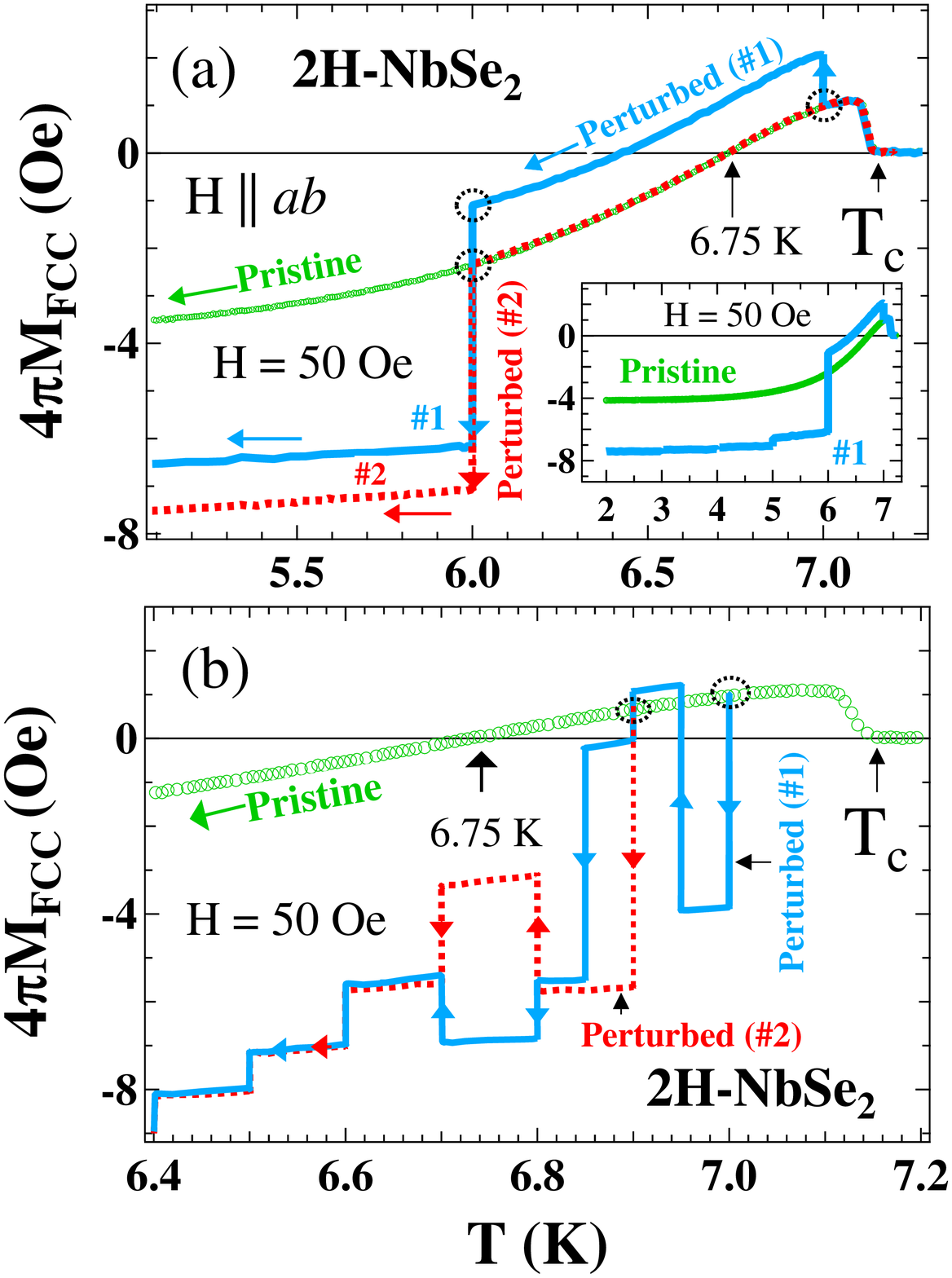}
\par\end{centering}

\begin{singlespace}
\protect\caption{(Color online) $M_{FCC}\thinspace(T)$ recorded at $H=50$\,Oe in
various runs perturbed at different temperatures. (a) Perturbed (\#1)
shows an enhanced paramagnetism at $7$\,K and an enhanced diamagnetism
at $6$\,K when compared with the pristine $M_{FCC}\thinspace(T)$.
In case of perturbed (\#2), an impulse at $6$\,K displays equal
change in magnetization as is seen in case of perturbed (\#1). Inset
shows pristine and perturbed (\#1) in the full temperature range.
(b) $M_{FCC}\thinspace(T)$ traced while an impulse is applied at
short temperature interval in two runs, perturbed (\#1) and perturbed
(\#2), leading to a characteristic path dependence above $6.75$\,K,
whereas below it, both follows the same path.}
\end{singlespace}
\end{figure}

\subsection{Path dependent $M(T)$ response}

We now demonstrate in Fig.~4, the impact of an ac field impulse applied
at various temperatures, during a given $M_{FCC}\thinspace(T)$ run
at $H=50$\,Oe. The pristine $M_{FCC}\thinspace(T)$ shown in Fig.~4
is the same as presented in Fig.~3. When an impulse was applied at
$T=7$\,K (encircled in Fig.~4(a)), an enhanced paramagnetic signal
was seen as depicted by the $M_{FCC}\thinspace(T)$ curve named \textquotedblleft perturbed
(\#1)\textquotedblright . This curve crosses over to the diamagnetic
region smoothly at $T\sim6.4$\,K, and thereafter remains less diamagnetic
(than the pristine $M_{FCC}\thinspace(T)$) until about $T=6$\,K,
echoing the \textquotedblleft perturbed (\#2)\textquotedblright{}
case of Fig.~3. At $T=6$\,K, an impulse was imposed again (see
the upper encircling done corresponding to $6$\,K in Fig.~4(a))
and the $M_{FCC}$ curve thereafter (i.e., perturbed (\#1)) becomes
more diamagnetic than the pristine one. The field-cooled curve (perturbed
(\#1)) follows a path shown by a continuous line below $6$\,K in
Fig.~4(a),which is different from its pristine counterpart. The saturated
$M_{FCC}\thinspace(T)$ value at $2$\,K corresponding to perturbed
(\#1) curve is found to be nearly twice the saturated $M_{FCC}\thinspace(T)$
in the pristine case, as is evident in the inset panel of Fig.~4(a).
In another run, we applied an impulse once again at $6$\,K, but
on the pristine field-cooled (FC) state (see another encircling done
at $6$\,K in Fig.~4(a)). Such an impulse treatment resulted in
the diamagnetic value of pristine $M_{FCC}\thinspace(T)$ to switch
over to a still larger diamagnetic value. The $M_{FCC}\thinspace(T)$
curve thereafter, i.e., perturbed (\#2) shown by the dashed line in
Fig.~4(a), can be seen to be even more diamagnetic than the other
two curves. We also found that after several iterations, that the
change in magnetization induced due to impulse treatments below about
$6.7$\,K is only of one kind, i.e., from a given diamagnetic value
to another diamagnetic value, unlike the behaviour noticed above $6.75$\,K
(i.e., in the PME region), where one could see switching from para/diamagnetic
to more or less para/diamagnetic value (see Fig.~3).

In Fig.~4(b), we show the $M_{FCC}\thinspace(T)$ curves at $H=50$\,Oe,
when an ac field impulse is consecutively imposed at short temperature
interval(s) in a given FC run. We first applied an impulse at $T=7$\,K
(encircled in Fig.~4(b)), and observed a switching of paramagnetic
$M_{FCC}$ of the pristine state into a diamagnetic value (perturbed
(\#1)) as shown by continuous line. This $M_{FCC}\thinspace(T)$ curve
(perturbed (\#1)) traces the record while cooling the sample further
down from $7$\,K to $6.95$\,K, where the application of another
impulse switches the diamagnetic value into a paramagnetic value,
which is even more paramagnetic than the pristine one. As the cooling
was progressed to $T=6.9$\,K, an impulse treatment there lead to
the change from paramagnetic $M_{FCC}$ value to near zero magnetization
value. This process of applying an ac impulse at short intervals was
continued, and the $M_{FCC}\thinspace(T)$ curve (\#1) displayed in
Fig.~4(b) was traced. In another field-cooling run, we applied the
impulse for the first time at $T=6.9$\,K, which resulted in switching
from paramagnetic to a diamagnetic value as shown by the dashed curve
(perturbed (\#2)) in Fig.~4(b). This curve was recorded while further
cooling, it also comprised impulse treatment(s) at several intermediate
temperatures (see steps in magnetization where the impulse was applied)
during the cool-down process.

An examination of $M_{FCC}\thinspace(T)$ data recorded in the two
runs, \textquotedblleft perturbed (\#1)\textquotedblright{} and \textquotedblleft perturbed
(\#2)\textquotedblright , in Fig.~4(b) shows a pronounced path-dependence
in $M_{FCC}\thinspace(T)$ at temperatures above about $6.7$\,K.
This suggests that the impact of an impulse above $6.7$\,K is completely
unpredictable: the system can display any magnetization value ranging
from paramagnetic to diamagnetic. The impulse seems to be acting as
a drive which brings out transition from a given (paramagnetic/diamagnetic)
state of the system into another and hence leads to a non-uniqueness
in $M_{FCC}\thinspace(T)$. On the other hand, the two $M_{FCC}\thinspace(T)$
runs (\textquotedblleft perturbed (\#1)\textquotedblright{} and \textquotedblleft perturbed
(\#2)\textquotedblright ) tend to overlap when an impulse was applied
below about $6.7$\,K. Below this temperature, the two curves respond
identically to any further impulse treatment and the magnetization
values remain diamagnetic.
\begin{figure}[t]
\begin{centering}
\includegraphics[scale=0.42]{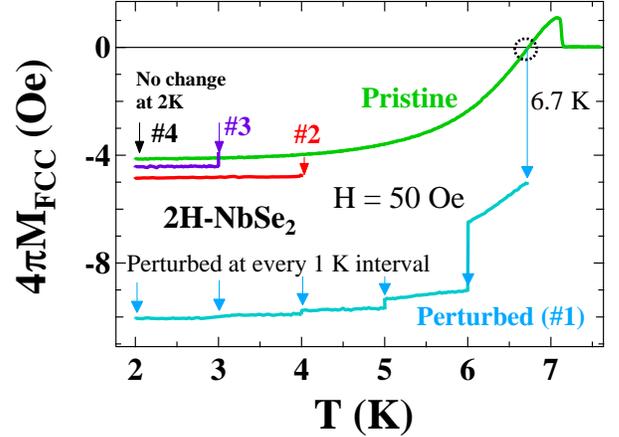}
\par\end{centering}

\begin{singlespace}
\protect\caption{(Color online) $M_{FCC}\thinspace(T)$ in pristine and perturbed cases
at $H=50$\,Oe. In perturbed (\#1), an impulse was first imposed
at $T=6.7$\,K (encircled), followed by another at $6$\,K, and
at every $1$\,K difference below $6$\,K. Saturated $M_{FCC}\thinspace(T)$
value of pristine and perturbed (\#1) are much different. In contrast,
the impulse applied at $4$\,K (perturbed (\#2)) and at 3K (perturbed
(\#3)) brings out minimal changes. No change is registered when the
impulse is applied at $2$\,K.}
\end{singlespace}
\end{figure}

To explore the highest saturated value of diamagnetic response, we
chose to successively apply an ac field impulse below $6.7$\,K at
every $1$\,K interval beginning from the pristine field-cooled (FC)
state created at $T\approx6.7$\,K as shown in Fig.~5. The arrows
in Fig.~5 indicate the temperatures at which the impulse was applied.
The first impulse applied at $6.7$\,K changes the magnetization
from a diamagnetic value (encircled) of pristine $M_{FCC}\thinspace(T)$
to a larger diamagnetic value (perturbed (\#1)). After lowering the
temperature to $6$\,K, another impulse drives the magnetization
into a still more diamagnetic state and so on. The saturated $4\pi M$
value for the perturbed (\#1) $M_{FCC}\thinspace(T)$ curve in Fig.~5
at $2$\,K happens to be about $-10$\,Oe, which is more than twice
the saturated $4\pi M$ value ($\approx-4$\,Oe) of the pristine
curve.

The commencement of application of an impulse at temperatures lower
than $4$\,K resulted in a small change in the magnetization. For
example, the changes in magnetization after the impulse treatment
at $T=4$\,K and $3$\,K in pristine $M_{FCC}\thinspace(T)$ runs
were comparatively smaller in the perturbed states \textquotedblleft \#2\textquotedblright{}
and \textquotedblleft \#3\textquotedblright , respectively. Eventually,
at $T=2$\,K (perturbed (\#4)), no change in pristine $M_{FCC}$
was witnessed after the application of an impulse at this temperature
(cf. Fig.~5). 
\begin{figure}[b]
\begin{centering}
\includegraphics[scale=0.32]{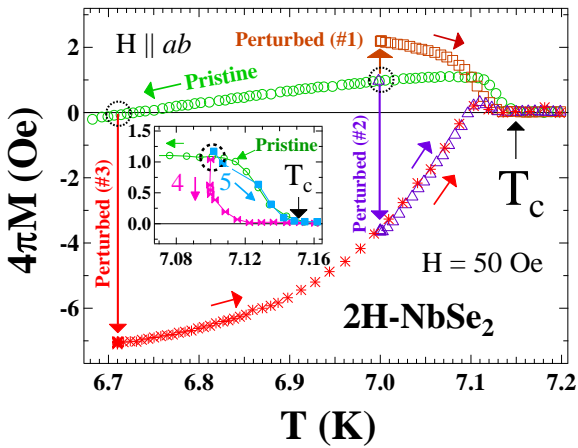}
\par\end{centering}

\begin{singlespace}
\protect\caption{(Color online) $M_{FCW}\thinspace(T)$ curves at $H=50$\,Oe traced
in five different warm-up runs after perturbation by an ac field impulse
(perturbed (\#1) to (\#5)). The pristine curve (open circles) which
was recorded during field-cooling in same field is also shown. Magnetization
data in the perturbed conditions (\#1 to \#5) were recorded while
warming the sample to higher $T$, after perturbing the pristine FC
state created at $T=7$\,K (perturbed (\#1) and perturbed (\#2)),
$T=6.71$\,K (perturbed (\#3)) and $T=7.1$\,K (inset shows perturbed
(\#4) and (\#5)). In all the five cases, the $M_{FCW}\thinspace(T)$
curves show a variety of paths that can be traversed while warming
after the perturbation.}
\end{singlespace}
\end{figure}

The results in Figs.~4 and 5 focus on $M_{FCC}\thinspace(T)$ data,
while subjecting the sample to perturbations at intermediate temperatures
during cool-down cycle. In Fig.~6, we show another set of magnetization
data, wherein we present the $M_{FCW}\thinspace(T)$ responses recorded
while sweeping the temperature up after an impulse treatment to (pristine)
FC states created at various temperatures. In the first run, after
getting an enhanced paramagnetism for impulse applied at $T=7$\,K
(encircled in main panel of Fig.~6), we recorded the magnetization
(perturbed (\#1)) while warming the sample towards the normal state.
The paramagnetic values for \textquotedblleft perturbed (\#1)\textquotedblright{}
fall smoothly as the temperature approaches $T_{c}$, however, this
curve does not merge with the pristine $M_{FCC}\thinspace(T)$ data.
Magnetization data in the \textquotedblleft perturbed (\#2)\textquotedblright{}
(open triangles) and \textquotedblleft perturbed (\#3)\textquotedblright{}
(stars) states showed a switching towards diamagnetic region after
the impulses were applied at $7$\,K and $6.71$\,K, respectively.
While warming-up, the $M_{FCW}$ curve for \textquotedblleft perturbed
(\#3)\textquotedblright{} merges with that of \textquotedblleft perturbed
(\#2)\textquotedblright{} at $T=7$\,K, and the two curves together
follow nearly the same path thereafter. It is a mere coincidence that
the first magnetization data after the impulse treatment at $7$\,K
in the case of perturbed (\#2) falls on the warm-up curve of perturbed
(\#3). Taking cue from the observations made in Fig.~3, the magnetization
after the impulse treatment at $7$\,K could have achieved any value
ranging from paramagnetic to diamagnetic. Some examples of path-dependent
$M_{FCW}\thinspace(T)$ responses in the close proximity of $T_{c}$
are displayed in an inset panel of Fig.~6. In the fourth run, we
applied the impulse at $T=7.10$\,K, and recorded the $M_{FCW}\thinspace(T)$
thereafter while warming up the sample (see $M_{FCW}$ for \textquotedblleft perturbed
(\#4)\textquotedblright{} in the inset of Fig.~6). This curve can
be seen to follow a path different from the pristine cool-down one,
as is apparent from the inset panel. When we once again applied the
impulse at $T=7.10$\,K i.e., the $5^{th}$ attempt, the $M_{FCW}(T)$
curve (perturbed (\#5)) nearly traced the path of the pristine $M_{FCC}$
curve. Thus, during the warm-up runs also, one can observe a variety
and path dependence in $M(T)$ response after creating different perturbed
states (see \textquotedblleft \#1\textquotedblright{} to \textquotedblleft \#5\textquotedblright{}
in Fig.~6).

\begin{figure}[t]
\begin{centering}
\includegraphics[scale=0.3]{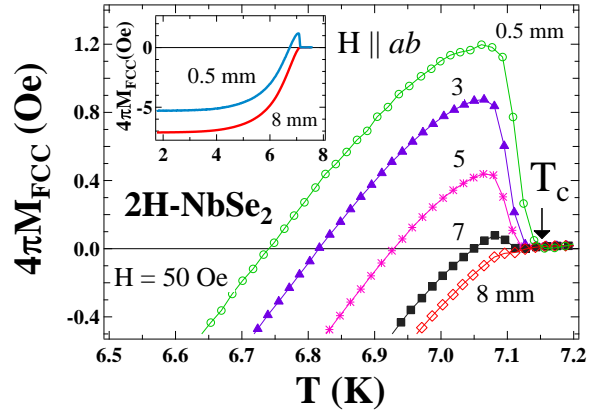}
\par\end{centering}

\begin{singlespace}
\protect\caption{(Color online) Portions of the $M_{FCC}\thinspace(T)$ curves recorded
in $H=50$\,Oe at various amplitudes of vibration in a VSM, as indicated.
The saturated values of $M_{FCC}$ at $T\sim1.8$\,K for two of the
curves, namely those recorded for 0.5\,mm and 8\,mm, can be seen
to be different in the inset panel.}
\end{singlespace}
\end{figure}

\subsection{Diversity in $M_{FCC}$ across the region of PME}

We have so far restricted our discussion to the effect of a perturbation
in the form of an ac field impulse on the FC magnetization response
(Figs.~3 to 6). It is also tempting to ask whether a multiplicity
in $M(T)$ similar to that observed in Figs.~3 to 6 can also be explored
by any other form of perturbation, say, via a change in the experimental
conditions. In this spirit, we show in the main panel of Fig.~7,
a variety in the $M_{FCC}\thinspace(T)$ curves obtained at a chosen
field of $50$\,Oe, when the amplitude of sample vibration in SVSM
was changed from 0.5\,mm to 8\,mm. The following features in Fig.~7
are noteworthy:
\begin{enumerate}
\item The $M_{FCC}\thinspace(T)$ curves at different amplitudes traverse
different paths, which is very striking. In particular, the saturated
values at the lowest temperature are significantly different during
different runs (data shown only for the amplitudes of 0.5\,mm and
8\,mm in the inset of Fig.~7). Such a difference in saturated $M_{FCC}\thinspace(T)$
is very similar to the observations made earlier for the $M_{FCC}(T)$
responses of the pristine and perturbed FC states (cf. inset panel
of Fig.~3).
\item The paramagnetic peak can be seen to have highest value during the
$M_{FCC}\thinspace(T)$ run with an amplitude of 0.5\,mm. The peak
height was observed to steadily decrease as the amplitude was progressively
increased from 0.5\,mm to 7\,mm. 
\item At an amplitude of 8\,mm, the $M_{FCC}\thinspace(T)$ values can
be seen to be diamagnetic in the entire temperature range $1.8$\,K\,$<T<T_{c}$
(cf. the curve for 8\,mm amplitude in the main panel and in the inset
panel of Fig.~7).
\end{enumerate}
Figure~7 thus illustrates a rich multiplicity in $M_{FCC}\thinspace(T)$
getting exposed by the change in the amplitude of sample vibration.
The similarity between the features emanating from $M_{FCC}\thinspace(T)$
curves of inset panel in Fig.~7 and those of inset panel in Fig.~3
is curious.

\begin{figure}[t]
\begin{centering}
\includegraphics[scale=0.35]{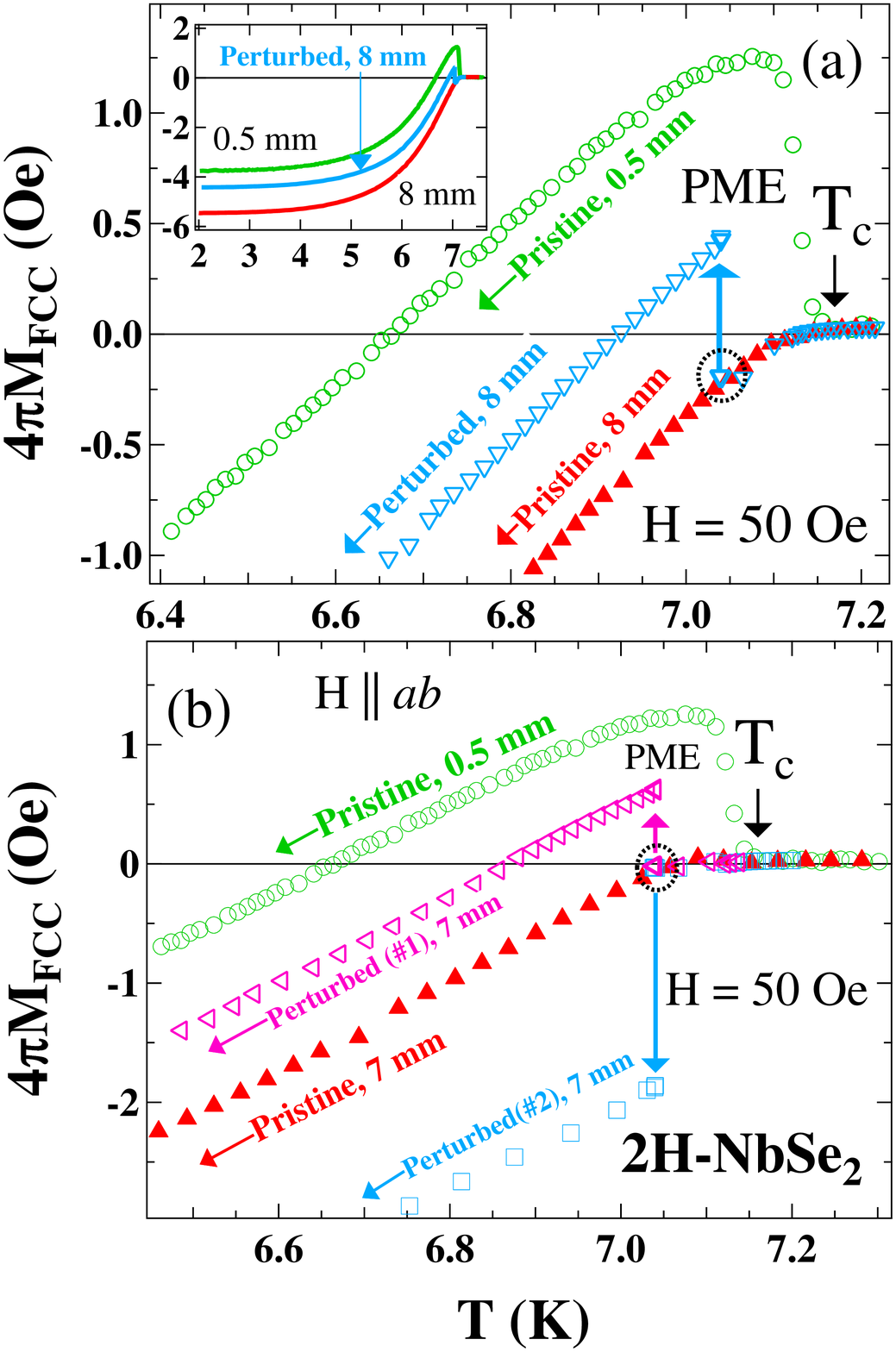}
\par\end{centering}

\begin{singlespace}
\protect\caption{(Color online) (a) $M_{FCC}(T)$ in $H=50$\,Oe, recorded at amplitude
of 0.5\,mm and 8\,mm, without imposing a perturbation during the
cool-down are termed as pristine curves. The impulse treatment at
$7.05$\,K switches a diamagnetic value of pristine $M_{FCC}$ (8\,mm)
into a paramagnetic value, and the (perturbed, 8\,mm) $M_{FCC}\thinspace(T)$
curve was traced on further cooling. The inset panel in Fig.~8(a)
displays the three curves in the range, $2$\,K\,$<T<7.5$\,K.
(b) $M_{FCC}\thinspace(T)$ at $H=50$\,Oe obtained at amplitudes
of 0.5\,mm and 7\,mm, respectively in pristine conditions, plotted
together with $M_{FCC}\thinspace(T)$ obtained at 7\,mm in the perturbed
situations (\#1 and \#2). In perturbed (\#1) state (open triangles),
the $M_{FCC}\thinspace(T)$ switches from near zero value to a paramagnetic
value whereas in other case (\#2), it switches into the diamagnetic
region (open squares).}
\end{singlespace}
\end{figure}
It was observed earlier in Fig.~4(b) that an impulse, when applied
above about $6.8$\,K, can switch a given diamagnetic value to a
paramagnetic value (see perturbed (\#1) in Fig.~4(b)). Along this
line, it will now be interesting to investigate whether such a change
can also be induced by an impulse in a $M_{FCC}\thinspace(T)$ response
at higher amplitude, say 8\,mm, which in pristine conditions remained
entirely diamagnetic below $T_{c}$ (in Fig.~7). To demonstrate this,
we applied an impulse on the pristine FC state created (at 8\,mm
amplitude) at $T=7.05$\,K in a field of $H=50$\,Oe as shown in
Fig.~8(a). For comparison, we have appended in Fig.~8 the pristine
$M_{FCC}\thinspace(T)$ curve at 0.5\,mm amplitude as well. It is
interesting to note that the diamagnetic value of the pristine $M_{FCC}\thinspace(T)$
curve for 8\,mm amplitude changes to a paramagnetic value after the
impulse treatment at $T=7.05$\,K. The $M_{FCC}\thinspace(T)$ curve
after perturbation (open triangles) was traced and found to be following
a different path as shown in the main panel of Fig.~8(a). The saturated
value of this curve (perturbed, 8\,mm) at $2$\,K approaches nearer
to that of pristine $M_{FCC}\thinspace(T)$ curve at $0.5$\,mm as
seen in the inset of Fig.~8(a). Similar investigations were carried
out for the pristine $M_{FCC}(T)$ curve at $7$\,mm amplitude as
shown in Fig.~8(b). Here, we performed two different runs (independently),
after the impact of an impulse at $T=7.05$\,K. In one of the runs
(perturbed (\#1)), a paramagnetic magnetization was induced at $7.05$\,K
and the $M_{FCC}\thinspace(T)$ thereafter follows a certain path
(open triangles), which is similar to that seen for 8\,mm amplitude
in the perturbed state (Fig. 8(a)). During the other run (perturbed
(\#2)), we observed that the magnetization data changes from near
zero value of the pristine $M_{FCC}$ for $7$\,mm to a large diamagnetic
value after the imposition of an impulse at $7.05$\,K, and follows
a path shown by open squares in Fig.~8(b). We draw an important inference
here that the occurrence of paramagnetic response is not specific
to a given amplitude of sample vibration, as it can be unearthed via
a perturbation at all amplitudes ranging from 0.5\,mm to 8\,mm.
\begin{figure}[t]
\begin{centering}
\includegraphics[scale=0.32]{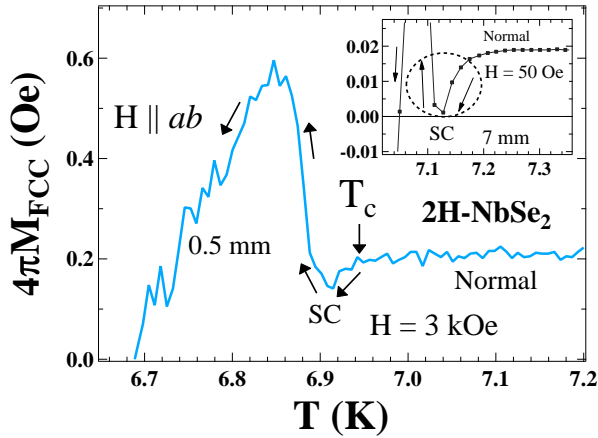}
\par\end{centering}

\begin{singlespace}
\protect\caption{(Color online) A portion of the $M_{FCC}\thinspace(T)$ curve at $H=3$\,kOe
recorded at an amplitude of 0.5\,mm showing some unusual undulations.
The inset panel shows a dip (see encircled portion) in $M_{FCC}\thinspace(T)$
curve at $H=50$\,Oe, just below the onset of the superconductivity
during run with $7$\,mm amplitude.}
\end{singlespace}
\end{figure}

We finally exemplify in the main panel of Fig.~9 an unusual feature,
viz., a characteristic oscillatory behaviour seen in $M_{FCC}\thinspace(T)$
response recorded at 0.5\,mm amplitude in a higher field, $H=3$\,kOe.
This curve exhibits some peculiar undulations which override the paramagnetic
magnetization signal below $T_{c}$. It appears as if the magnetization
shows a tendency to switch between the paramagnetic and diamagnetic
responses while field-cooling. Such undulations echo similar oscillatory
features reported earlier at lower field values ($<14$\,Oe) in another
low $T_{c}$ superconductor, viz., Ca$_{3}$Rh$_{4}$Sn$_{13}$ by
Kulkarni \textit{et al.} \cite{key-24}. It may also be added here
that the oscillatory behaviour in $M_{FCC}\thinspace(T)$ is not merely
restricted to lower amplitudes and higher fields, but, the same were
also discernible at higher amplitudes and lower fields in 2H-NbSe$_{2}$.
At 7\,mm amplitude and in a field of $H=50$\,Oe, the $M_{FCC}\thinspace(T)$
response was observed to have an oscillatory character in close proximity
of $T_{c}$ as can be seen in an expanded portion of the $M_{FCC}\thinspace(T)$
in an inset panel of Fig.~9. This curve first shows a pronounced
dip (encircled in the inset of Fig.~9), which can be taken as the
onset of diamagnetism. Following this dip, there occurs a sharp upturn
on further lowering of temperature. The $M_{FCC}\thinspace(T)$ values
eventually cross over to negative values on lowering the temperature
below $T\approx7.05$\,K.

\section{Discussion}

We have come across some amazing metastability effects in magnetization
response on field-cooling a single-crystal of 2H-NbSe$_{2}$, viz.,
occurrence of both paramagnetic and diamagnetic response in a temperature
window just below $T_{c}$, multi-valuedness in $M_{FCC\thinspace}(T)$
at a given ($H$,$T$) value, oscillatory behaviour in a given $M_{FCC\thinspace}(T)$
curve, unpredictable switching of $M_{FCC\thinspace}(T)$ between
diamagnetic and paramagnetic values due to an intervention by an external
perturbation, etc. The mechanism and origin behind the said features
in 2H-NbSe$_{2}$ cannot be comprehended on the basis of existing
theoretical treatments in the literature. The following descriptions
form the basis of this assertion:

1. The occurrence of (anomalous) paramagnetic magnetization in conventional
bulk superconductors \cite{key-13} had lead to several proposals,
amongst which the flux-trapping model by Koshelev and Larkin \cite{key-31}
received significant attention. This model treats the situation wherein
the magnetic flux gets trapped while field-cooling within a (macroscopic
or mesoscopic) superconducting specimen due to an inhomogenous cooling
\cite{key-31}. While a (diamagnetic) current flows on the surface
to shield an external field, associated with the trapped flux inside
the specimen, there occurs a (paramagnetic) current that circulates
in the interior of a superconducting sample \cite{key-31}. Paramagnetic
magnetization response can result due to an unbalancing effect of
magnetization responses arising from the above stated two currents.
The inhomogeneity in field-cooling, can have a variety, which in turn,
can lead to a variety in trapped flux, thereby, resulting in multitude
of magnetization responses during field-cooling experiments. Such
a possibility could rationalize the observation of variety in $M_{FCC\thinspace}(T)$
response, as reflected in Fig.~7 in the form of vibration amplitude\,-\,dependent
$M_{FCC\thinspace}(T)$ data at a given field value. Carrying this
argument further, the data in Fig.~7 would imply that the inhomogeneity
in cooling may be lesser at higher value of amplitude of vibration
(see $M_{FCC\thinspace}(T)$ curves at 7\,mm and 8\, mm amplitude
in Figs.~7 and 8), leading to insignificant flux trapping (absence
of paramagnetic signal). However, even in case of $M_{FCC\thinspace}(T)$
data recorded at higher amplitudes (say 8\,mm), that had remained
entirely diamagnetic below $T_{c}$, a diamagnetic $M_{FCC\thinspace}(T)$
value at a given temperature can be switched into a paramagnetic one,
when the sample is perturbed by an impulse of an ac field (cf. Fig.~8(a)).
From the perspective of flux-trapping model \cite{key-31}, this would
imply that the impulse treatment can drive the system from a state
of insignificant trapped-flux (diamagnetic) to a state of large trapped-flux
(paramagnetic), which sounds to be unfeasible. Along similar lines,
it can be argued that the occurrence of a diamagnetic dip just below
$T_{c}$ (insignificant trapped-flux) followed by a paramagnetic peak
(large trapped-flux) in $M_{FCC\thinspace}(T)$ (cf. Fig.~9) cannot
be rationalized by this model. Also, the switching tendency of magnetization
between diamagnetic and paramagnetic values identified via the characteristics
undulations (oscillatory behaviour) in a given $M_{FCC\thinspace}(T)$
run (Fig.~9) needs comprehension which is beyond the scope of flux-trapping
model. We further noted two different effects of an impulse in two
independent runs on the $M_{FCC\thinspace}(T)$ response obtained
at a higher amplitude of 7\,mm (Fig.~8(b)). In one of the instances
(perturbed (\#1)), the impulse switches a diamagnetic (pristine) $M_{FCC\thinspace}(T)$
value into a paramagnetic one while in the other (independent) run
(\#2), it switches the same diamagnetic $M_{FCC\thinspace}(T)$ value
into a larger diamagnetic value. Clearly, this suggests that the effects
of an impulse in the present study are completely \textit{random,}
whereas the Koshelev and Larkin\textquoteright s mathematical treatment
\cite{key-31} to a given field and current distribution due to a
trapped flux ought to be \textit{predictable}. We therefore rule out
that this model can explain the metastability observed in $M_{FCC\thinspace}(T)$
response of 2H-NbSe$_{2}$.

2. To understand the thermomagnetic history dependence observed in
the magnetization data of a type-II superconductor, Clem and Hao \cite{key-42}
had constructed a model which relies on the Bean\textquoteright s
Critical State framework \cite{key-43}. Although this model \cite{key-42}
yields the flux-density profiles that can get set up in a superconductor
during different protocols usually studied by experimentalists, viz.,
zero field-cooled, field-cooled cool-down and field-cooled warm-up
modes, it cannot account for paramagnetic magnetization in the superconducting
specimen. However, we noted that the effect of an impulse applied
below $T=6.7$\,K does not lead to any paramagnetic magnetization
but an enhanced diamagnetism (cf. Fig.~5). The multi-valuedeness
in (diamagnetic) $M_{FCC\thinspace}(T)$ data at low temperatures
($T<6.7$\,K, cf. inset panel of Fig.~3, 4(a) and 7 and 8(a) and
Fig. 5) can be related in an oblique way to a result that emanated
using the model of Clem and Hao \cite{key-42}. We recall that the
flux expulsion by a pinned-superconductor when it is field-cooled
through its $T_{c}$ is never complete \cite{key-42} and therefore
the field-cooled magnetization generally does not conform to an equilibrium
magnetization. The model by Clem and Hao \cite{key-42} indeed shows
that the $M_{FCC\thinspace}(T)$ curve at a constant field depends
on the pinning strength; a more (less) diamagnetic state amounts to
less (more) strongly pinned vortex configuration (see Fig.~5 in Ref.~\cite{key-42}).
We observe a diversity in $M_{FCC\thinspace}(T)$ data (see, for example,
inset of Fig.~3) that closely resembles the pinning-dependent (multi-valued)
$M_{FCC\thinspace}(T)$ curves numerically calculated by Clem and
Hao (see Fig.~5 in Ref.~\cite{key-42}), however, the multiplicity
seen in present case is in the same 2H-NbSe$_{2}$ crystal, i.e.,
without changing the amount of quenched-disorder (pinning strength).
Comparing the two pictures, viz., the $M_{FCC\thinspace}(T)$ response
of 2H-NbSe$_{2}$ (an inset of Fig.~3) and that shown in Fig.~5
in Ref.~\cite{key-42}, one may naively surmise that the non-uniqueness
in $M_{FCC\thinspace}(T)$ in 2H-NbSe$_{2}$ could be due to the possibility
of coexistence of several (metastable) vortex configurations with
varying pinning strength. The impulse treatment or amplitude-variation
can lead to the switching of the system from one configuration to
another, which is reflected by the diversity in $M_{FCC\thinspace}(T)$
as shown in inset panels of Figs.~3 and 7. In a given field-cooled
run, the $M_{FCC\thinspace}(T)$ values get saturated below a certain
temperature, $T_{c1}$, at which the applied field just equals the
lower critical field ($H_{c1}(T)$). Below $T_{c1}$, i.e., in the
Meissner phase, the applied field remains smaller than the $H_{c1}(T)$
value and the application of an ac field impulse ($h_{ac}$) superimposed
on the (static) applied field ($H\pm h_{ac}$) is hardly expected
to produce any significant change in $M_{FCC\thinspace}(T)$. This
is apparent from the minimal changes seen in $M_{FCC\thinspace}(T)$
below $T=5$\,K for the perturbed cases (\#2), (\#3) and (\#4) in
Fig.~5.

3. 2H-NbSe$_{2}$ system has for long remained a favourite compound
to explore novel notions in vortex state studies. In particular, this
compound has been very widely investigated (see Ref.~\cite{key-44}
and references therein) for order-disorder transitions as fingerprinted
via anomalous variations in field/temperature dependencies in critical
current density, viz., peak effect (PE) phenomenon/second magnetization
peak (SMP) anomaly in magnetization hysteresis loops. Due to supercooling/superheating
effects \cite{key-45} that can occur across these transitions, the
magnetization response is generally found to be history-dependent
\cite{key-44}. However, we would like to emphasize here that the
present results pertaining to metastability effects in 2H-NbSe$_{2}$
have no correlations with the metastability and thermomagnetic history-dependence
in critical current density values seen across the PE/SMP transitions.
Firstly, most of the magnetization data presented here have been recorded
at very low fields (i.e., $H=50$\,Oe). We have checked that at this
field value, the present crystal does not display either PE in temperature-dependent
isofield scans or SMP transition in isothermal scans at any temperature.
Secondly, the history-dependent magnetization across the PE/SMP has
never been reported to generate paramagnetic magnetization. 

4. Clearly, all the novel proposals \cite{key-4,key-8,key-9,key-27,key-28,key-29,key-30}
that have been put forward to understand the origin of paramagnetic
magnetization in high\,-\,$T_{c}$ superconductors are very difficult
to visualize in the present context which involves a conventional
low\,-\,$T_{c}$ superconductor. Even the controlled manipulation
of paramagnetic magnetization demonstrated recently \cite{key-26}
in Pb-Co nanocomposites involved the change in the orientation of
magnetic moments of Co nanoparticles unlike the present situation
wherein we have observed in a non magnetic superconductor, 2H-NbSe$_{2}$,
an unpredictable switching of paramagnetic magnetization into diamagnetic
and vice versa.

5. We rule out that the present results are a consequence of any specific
disorder present in the anisotropic 2H-NbSe$_{2}$ crystal as there
have been few other reports of similar metastability effects (including
oscillatory magnetization behaviour) in other isotropic low\,-\,$T_{c}$
superconductors, viz., single-crystals of Ca$_{3}$Rh$_{4}$Sn$_{13}$
\cite{key-24} and Nb (in the form of a sphere) \cite{key-18}. An
advantage with a spherical (Nb) crystal \cite{key-18} is that the
demagnetization factor remains the same irrespective of the sample
orientation with respect to field and, hence, yields identical results
in all orientations. In these studies \cite{key-18,key-24}, however,
the effects of perturbation on $M_{FCC\thinspace}(T)$ response were
not explored. The novel consequences of perturbation on the $M_{FCC\thinspace}(T)$
response in the present study have lead to exposure of a rich multiplicity
in \textit{FC} magnetization and random switching tendency of $M_{FCC\thinspace}(T)$
between paramagnetic and diamagnetic values, which is indeed an advancement
over the results reported in Ref.~\cite{key-18,key-24}.

We now focus our attention onto some theoretical works \cite{key-32,key-33,key-36,key-37,key-38,key-39,key-40}
which attempt to account for the occurrence of PME in mesoscopic-sized
samples. Based on the Ginzburg-Landau theory, the theoretical pictures
framed exclusively for mesoscopic superconductors foresee the occurrence
of giant vortex states with multiple flux quanta ($L>1$) at the onset
of surface superconductivity \cite{key-34}. Each multi-quantum state
has its own field domain of existence; at higher field end of each
domain, the response is diamagnetic, which crosses over to the paramagnetic
values as the field decreases (see Fig.~2(b) in Ref.~\cite{key-33},
Figs.~16, 17 and 23 in Ref.~\cite{key-40}). It is also possible
that the temperature (or field) sweeping in a given $M(T)$ (or $M$--$H$)
run is sufficient enough to drive the system away from the field/temperature
domain of a certain configuration ($L$), which may induce a forced
transition \cite{key-33,key-40} to the nearest minimum energy configuration
(with different $L$ value). An examination of the $M$--$H$ plots
for different multi-quanta states ($L>1$) along with the corresponding
free energy vs field curves drawn for a specific circumstance in Figs.~2
and 3 of Ref.~\cite{key-33} tells us that if the transitions occur
between $L$ states such that the configuration pertaining to a certain
$L$ value with the lowest free energy prevails, there would be no
metastability effects in the magnetization response and the magnetization
values as a function of temperature (in isofield scans) would be diamagnetic
all through. However, if metastability is permissible and transitions
happen between different metastable $L$ states, one can encounter
all sorts of possibilities, i.e., the magnetization values can change
from a given paramagnetic/diamagnetic value to a higher or a lower
value, and it can also transit from positive to negative values or
vice versa.

To our knowledge, the above mentioned (and other related) studies
\cite{key-32,key-33,key-36,key-37,key-38,key-39,key-40} describing
the magnetization behaviour of multi-quanta states ($L>1$) in mesoscopic
samples are the only examples which bear some resemblance with our
findings. We draw here some parallels between these theoretical findings
\cite{key-32,key-33,key-36,key-37,key-38,key-39,key-40} and our experimental
results in a single crystal of 2H-NbSe$_{2}$, to which the latter
results \cite{key-32,key-33,key-36,key-37,key-38,key-39,key-40} prima
facie do not apply. As mentioned above, each multi-quantum state ($L>1$)
has a certain domain of its existence wherein the magnetization response
can be either diamagnetic or paramagnetic depending upon the external
field \cite{key-33}. Similarly, in 2H-NbSe$_{2}$, there exists a
temperature window ($6.7$\,K\,$<T<T_{c}$), where a perturbation
applied to the system (at a given ($H$,$T$) value) can randomly
yield the magnetization which is either paramagnetic or diamagnetic
(see Figs.~3, 4(b), 6 and 8). The switching of $M_{FCC\thinspace}(T)$
due to a perturbation leads to a rich diversity in magnetization response
in 2H-NbSe$_{2}$ (cf. Figs. 3 to 6 and 8), very similar to characteristics
of magnetization behaviour anticipated during the transformation amongst
various multi-quanta ($L>1$) states \cite{key-33,key-40}. Additionally,
the oscillatory behavior seen in $M_{FCC\thinspace}(T)$ curve (Fig.~9)
mimics the magnetization response predicted \cite{key-33,key-40}
for the transformations amongst multi-quanta states ($L>1$) during
the temperature-sweep. In an earlier unrelated study, the scanning
tunneling microscopy technique employed by Karapetrov \textit{et al.}
\cite{key-46}, enabled them to directly observe a coexistence of
multi-quanta vortex states and Abrikosov lattice in an array of submicroscopic
metal (gold) island embedded in a single crystal of 2H-NbSe$_{2}$.
The simultaneous existence of multi-quanta state of different vorticities
is thus not an unrealistic possibility.

\textit{In spite of a strong resemblance of our results with the characteristics
of multi-quanta states }\cite{key-32,key-33,key-36,key-37,key-38,key-39,key-40}\textit{,
the possibility of occurrence of multi-quanta states in the sample
investigated here remains a speculation considering the larger size
of this (macroscopic) sample. The effect of multi-vorticity in such
a macroscopic sample is an exotic proposition. Therefore, there is
currently a necessity to look for a more sophisticated model to fully
understand the new findings in 2H-NbSe$_{2}$ as well as earlier results
in Ca$_{3}$Rh$_{4}$Sn$_{13}$ \cite{key-24} and Nb \cite{key-18}.}

In the end, we rule out the possibility of any experimental artefact
arising due to change in amplitude of vibration considering the following
important points: (i) The factor which could be affected most by changing
the sample vibration is the magnetic field-inhomogeneity which, in
the present case, is found to be negligible ( $\sim10^{-2}$\,\LyXThinSpace Oe).
(ii) We did not register any change in the magnetization on varying
the amplitude at a given ($H$,$T$) value in the case of a standard
Indium sample. (iii) We have also shown that the PME like attribute
in 2H-NbSe$_{2}$ is not limited to only smaller amplitudes but the
same can also be induced at higher amplitudes (8\,mm), after the
impulse treatment.

\section{Conclusion}

We have investigated in detail a single crystal specimen of 2H-NbSe$_{2}$
via dc magnetization measurements. The new revelations in this compound
correspond to the observation of an anomalous paramagnetic (PME like)
signal(s) below $T_{c}$ and some peculiar consequences of applying
a perturbation at various temperatures on a given field-cooled state.
Based on the effects of perturbation, two distinct temperature intervals
have been identified. Across a certain temperature range lying just
below $T_{c}$, the PME\,-\,like signal can be manipulated by applying
an external perturbation, which unpredictably switches the $M_{FCC\thinspace}(T)$
response from a given para/diamagnetic value into a different para/diamagnetic
value. On the other hand, when the system is perturbed away from $T_{c}$
(below about $T=6.7$\,K), the effect of perturbation is of one kind,
i.e., it only enhances the diamagnetism such that no paramagnetic
signal is seen. The saturated (diamagnetic) $M_{FCC\thinspace}(T)$
value at the lowest temperature is found to be influenced by the effect
of perturbation applied closer to the region of PME. The said features
in $M_{FCC\thinspace}(T)$ data in a single crystal of 2H-NbSe$_{2}$
coincidentally bear similarities with the magnetization response predicted
\cite{key-32,key-33,key-36,key-37,key-38,key-39,key-40} for mesoscopic
samples, wherein, a possibility of nucleation of multi-quanta states
($L>1$) have been discussed in the literature. We believe the present
results in a bulk 2H-NbSe$_{2}$ crystal cannot be explained by any
of the existing mathematical treatments based on prescriptions of
Bean's Critical State model \cite{key-43} reported in the literature
and call for a new theoretical framework. Though, we have presented
experimental results pertaining to a given orientation of magnetic
field vis. a vis. crystallographic direction ($H$\,||\,$ab$) of
hexagonal 2H-NbSe$_{2}$, we have also carried out magnetization measurements
in more often studied orientation, viz., $H$\,||\,$c$, in 2H-NbSe$_{2}$
and obtained nearly identical features (data not included in the present
report). We may therefore state that assertions made in this report
are not specific to the anisotropic nature of the intrinsic superconducting
parameters of 2H-NbSe$_{2}$.

\noindent \textbf{\large{}Acknowledgments}{\large \par}

\noindent We would like to acknowledge Ulhas Vaidya from Tata Institute
of Fundamental Research (TIFR), Mumbai, India for sharing his results
in spherical Nb single crystal. One of us (Santosh Kumar) wishes to
thank Ulhas Vaidya for his help and assistance in the use of SVSM
system in TIFR in the initial phase of this work. Santosh Kumar would
also like to acknowledge the Council of Scientific and Industrial
Research, India for the grant of the Senior Research Fellowship. The
work on crystal growth and its characterization at the University
of Warwick was supported by EPSRC, UK under Grant EP/I007210/1. We
are also grateful to E. Zeldov and S.S. Banerjee for fruitful discussions.

\end{document}